# Developing a Machine-Learning Interatomic Potential for Non-Covalent Interactions in Proteins

*Lejia Zeng[1,2], Xintong Zhang[1,2], Yuchan Pei[1,2], Lifeng Zhao[2], Lan Hua[2], Jincai Yang[2], Niu Huang[1,2],\**

[1]Tsinghua Institute of Multidisciplinary Biomedical Research, Tsinghua University, Beijing 102206, China

[2]National Institute of Biological Sciences, 7 Science Park Road, Zhongguancun Life Science Park, Beijing 102206, China

KEYWORDS: Machine Learning Interatomic Potential (MLIP), Non-covalent Interactions (NCI), Active Learning (AL), Quantum Mechanics (QM), Force Field (FF)

\* To whom correspondence should be addressed. N. H. (Email) huangniu@nibs.ac.cn (Phone) 86-10-80720645 (Fax) 86-10-80720813

ABSTRACT

Machine learning interatomic potentials (MLIPs) enable efficient modeling of molecular interactions with quantum mechanical (QM) accuracy. However, constructing robust and representative training datasets that capture subtle, system-specific interaction motifs remains challenging. We introduce PANIP (PAirwise Non-covalent Interaction Potential), an ensemble MLIP model built upon the NequIP framework and trained on non-covalent interactions (NCIs) between protein-derived fragments. PANIP is trained using an automated multi-fidelity active learning (MFAL) workflow, in which a representative training subset—termed PDB-FRAGID (PDB Fragment Interaction Dataset), was distilled from an otherwise prohibitively large pool of fragment dimers extracted from the Protein Data Bank (PDB). PANIP retains ωB97X-D3BJ/def2-TZVPP-level accuracy and achieves mean absolute errors below 0.2 kcal/mol on out-of-distribution systems, demonstrating excellent transferability across diverse NCI motifs. Compared to the widely used ANI-2x potential, PANIP delivers substantially lower errors, particularly for charged and strongly interacting dimers. Coupled with a fragmentation-based energy decomposition scheme, PANIP estimates protein–ligand binding energies at near force-field computational cost yet QM-level accuracy, enabling its use as a fragment-based scoring function that rivals specialized docking scoring functions.

INTRODUCTION

The accurate modeling of non-covalent interactions (NCIs) within biomolecules at quantum-mechanical (QM) accuracy remains computationally prohibitive for large systems. The integration of QM methods with machine learning (ML) enables the efficient exploration of complex molecular systems[1–3]. Machine learning interatomic potentials (MLIPs) reproduce QM potential energy surfaces (PESs) with high fidelity, often achieving QM-level accuracy at a fraction of the computational cost for small- and medium-sized systems. These developments have opened new avenues for efficiently exploring complex conformational and interaction landscapes in chemistry and biology[4–7].

However, the reliability and transferability of MLIPs depend critically on the quality, diversity, and balance of the underlying training data. Many existing datasets and MLIP frameworks undersample chemical and conformational spaces[8–14], emphasize a limited set of molecular species, or are tuned to specific classes of systems, which restricts their applicability to new interaction motifs and environments[3,5,15–18]. Recent efforts have introduced large-scale, chemically diverse datasets and "universal" models that span broad chemical spaces[19,20]. However, these primarily emphasize universality, and when it comes to particular and subtle interaction motifs crucial for biological systems, important regions may remain underexplored. In particular, NCIs in proteins involve a rich variety of hydrogen bonding, electrostatic, dispersion, cation–π, and sulfur-containing interactions that are strongly context-dependent and sensitive to local geometry. Capturing these motifs at QM accuracy in a manner that generalizes across diverse protein environments remains a major challenge.

Generating new QM datasets tailored to specific systems or interactions is both labor-intensive and computationally expensive, sometimes negating the gains provided by ML acceleration. Moreover, when such system-specific datasets are integrated with existing large-scale datasets, the amount of additional training data should be carefully controlled to avoid excessive training burden while still filling critical gaps in chemical and conformational coverage. A central challenge is therefore to construct optimized, broadly representative training datasets that minimize redundant QM calculations while maximizing coverage of relevant NCIs. Achieving this goal requires intelligent data selection and model training workflows.

The Protein Data Bank (PDB)[21] provides a rich, experimentally grounded source of NCI geometries within biologically functional contexts[22–25]. Unlike synthetic datasets, PDB-derived geometries reflect realistic chemical diversity, spatial complexity, and thermodynamic relevance[22,23]. This makes the PDB an attractive foundation for building MLIPs that target protein NCIs and are designed to generalize across biologically relevant interaction types, including sidechain-sidechain, sidechain-backbone, and protein-water contacts. At the same time, PDB data exhibits inherent biases, for example, over-representation of certain interactions, resolution limitations, and crystallographic artifacts, which must be carefully considered when constructing datasets and training models.

Leveraging the PDB at scale for MLIP training presents two main challenges. First, generating accurate QM reference energies for the hundreds of millions of possible fragment geometries is computationally prohibitive, even with efficient density functional theory (DFT) methods. Second, the distribution of interaction types in the PDB is highly imbalanced: common interactions such as conventional hydrogen bonds, are heavily over-represented, while rarer but functionally important interactions, including certain cation-π, sulfur-aromatic, or ionic contacts, occur relatively infrequently. Naively labeling and training on all available dimers would not only be infeasible, but would also risk biasing model learning. Moreover, processing and integrating such a vast structural and energetic dataset during model training requires substantial computational resources and careful data-management strategies.

To address these challenges, a multi-fidelity active learning (MFAL)[26,27] workflow is employed to construct an efficient, representative dataset and to train an MLIP tailored to protein NCIs. This hierarchical approach integrates low-cost and high-level QM calculations: the composite density functional method r²SCAN-3c[28] serves as a low-fidelity oracle for large-scale energetic screening, while a machine learning surrogate identifies a representative and diverse subset for high-level refinement at the ωB97X-D3BJ/def2-TZVPP level[29,30]. This process yields the PDB Fragment Interaction Dataset (PDB-FRAGID) — a condensed yet representative subset containing only 8.7% of the original pool. Trained on PDB-FRAGID, the resulting ensemble model PANIP achieves accuracy comparable to ωB97X-D3BJ/def2-TZVPP calculations. Built on the Neural Equivariant Interatomic Potentials (NequIP) architecture[31], PANIP preserves spatial symmetries and accurately captures atomic environments. The model was rigorously validated against multiple benchmark sets, including low-energy dimers, optimized dimers, geometries from the Cambridge Structural Database (CSD)[32], and non-equilibrium conformations generated by random sampling. Using this framework, we systematically characterized NCI patterns across the entire PDB, with particular emphasis on underexplored sulfur-driven interactions. Finally, we demonstrate PANIP's potential utility by applying it as a fragment-based[33] scoring function for protein-ligand binding prediction in benchmark systems.

RESULTS

**PDB-FRAGID Dataset Construction.** A major challenge in developing MLIPs is constructing training sets that simultaneously provide broad coverage of chemically diverse NCIs and remain computationally tractable for high-level QM labeling. To address this, an iterative data selection strategy using MFAL was used to reduce the initial 36.3 million PDB-derived dimers into a compact, information-rich dataset, PDB-FRAGID. Using r²SCAN-3c as a low-fidelity oracle, interaction energies were computed for all dimers, and a NequIP-based surrogate model was iteratively refined to identify high-error dimers for inclusion. This process distilled the dataset to approximately 3.15 million dimers (8.7% of the original pool, Figure 1, Table S1), while preserving coverage across 17 fragment types and 153 dimer combinations and prioritizing chemically challenging systems such as charged and polar fragments (Figure 1). Although r²SCAN-3c provides a cost-effective baseline and correlates strongly with ωB97X-D3BJ (Figure S8), it shows systematic deviations for certain noncovalent interactions, including π-stacking and ionic contacts (RMSE=0.423 kcal/mol). Therefore, ωB97X-D3BJ/def2-TZVPP calculations were used to generate the final energy labels.

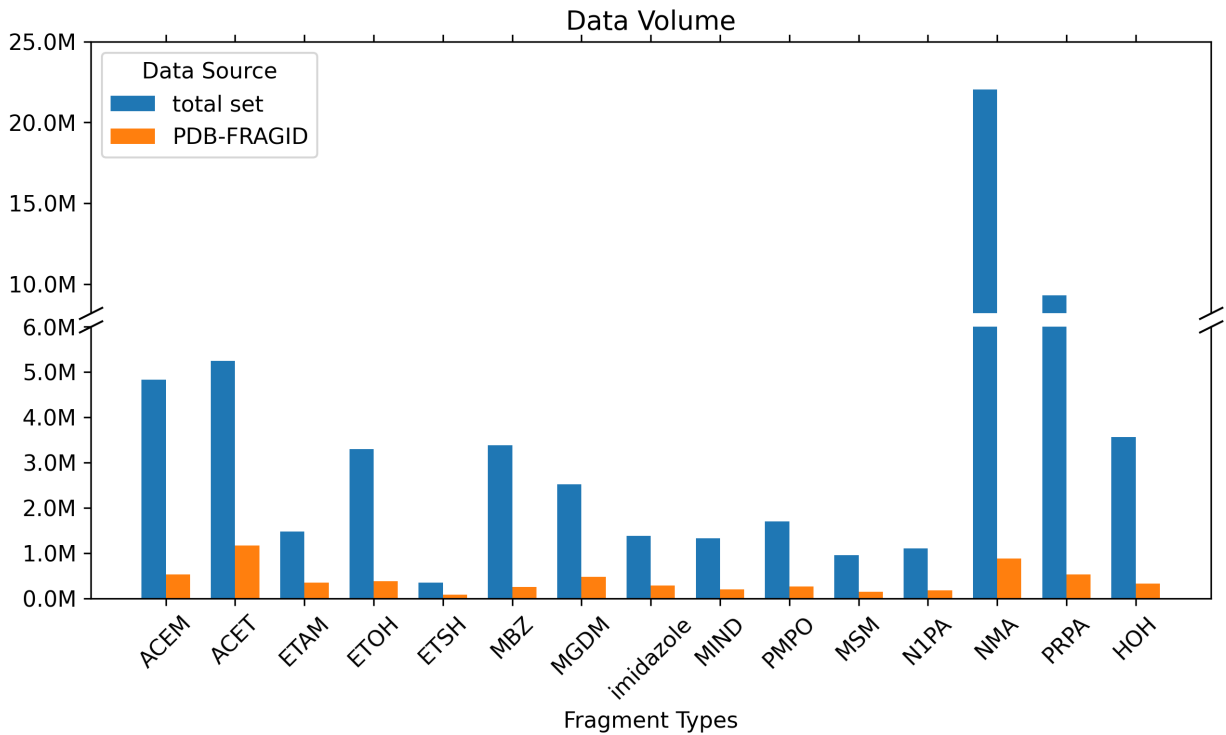

**Figure 1.** Data Volume for each fragment.

**Model Performance on Benchmark Sets.** PANIP was evaluated on four benchmark sets to assess accuracy, robustness, and transferability: low-energy dimers, optimized low-energy dimers, CSD-derived geometries, and non-equilibrium conformations generated by biased random sampling (Figure 2). For the PDB-derived low-energy dimers, PANIP achieved a mean absolute error (MAE) of 0.09 kcal/mol, a root mean squared error (RMSE) of 0.163 kcal/mol, and an $R^2$=0.999 relative to ωB97X-D3BJ/def2-TZVPP, indicating excellent reproduction of equilibrium NCI energies in the training domain (Figure 2A). After geometry optimization, the MAE and RMSE increased to 0.547 and 1.207 kcal/mol, respectively, but remained within chemically acceptable limits, indicating that the model can capture the energetic landscape near the equilibrium structures (Figure 2B).

On the CSD-derived benchmark set, which probes transferability to small-molecule crystal environments distinct from proteins, PANIP maintained strong performance (MAE 0.171 kcal/mol, RMSE 0.507 kcal/mol, $R^2$=0.999; Figure 2C), underscoring its generalization beyond PDB geometries. For 15,300 randomly sampled conformations, PANIP achieved an overall MAE of 0.448 kcal/mol, RMSE of 1.372 kcal/mol, and $R^2$=0.996. When restricted to physically plausible dimers (8,572 entries with interaction energy < 0 kcal/mol), the MAE and RMSE improved to 0.195 and 0.363 kcal/mol, respectively, with $R^2$=0.999 (Figure S4a). The subset with interaction energies above 0 kcal/mol (6,728 entries) showed a relatively higher error (RMSE 0.882 kcal/mol; Figure S4b). High-energy outliers (> 100 kcal/mol), representing less than 5% of the dataset, were systematically underestimated due to severe steric clashes and limited training analogs (e.g., ETSH dimers; Figure 1), but these geometries are rare in realistic biomolecular applications.

Computationally, PANIP delivers more than two orders-of-magnitude speedups over direct QM calculations. For example, evaluating 15,300 randomly generated samples on a single CPU core, ωB97X-D3BJ/def2-TZVPP calculations would require 463 days and 11 hours, whereas PNAIP predictions completed in just 6 hours and 11 minutes, reflecting near-linear scaling compared to the formal $N^4$ scaling of hybrid DFT methods (cubic with modern RI/RIJCOSX implementations; Table S2)[29].

Considering that NCIs are diverse and highly sensitive to subtle structural variations, even minor conformational changes can lead to pronounced energy differences[34,35]. The model's success in predicting unseen conformations underscores the sufficiency of PDB-derived dimers in sampling conformational space, enabling reliable generalization—a critical challenge in developing MLIPs. This emphasizes the importance of representative training sets for capturing diverse conformational and interaction landscapes. Such capabilities are particularly vital for applications in drug discovery and protein engineering, where accurate prediction of NCIs is essential.

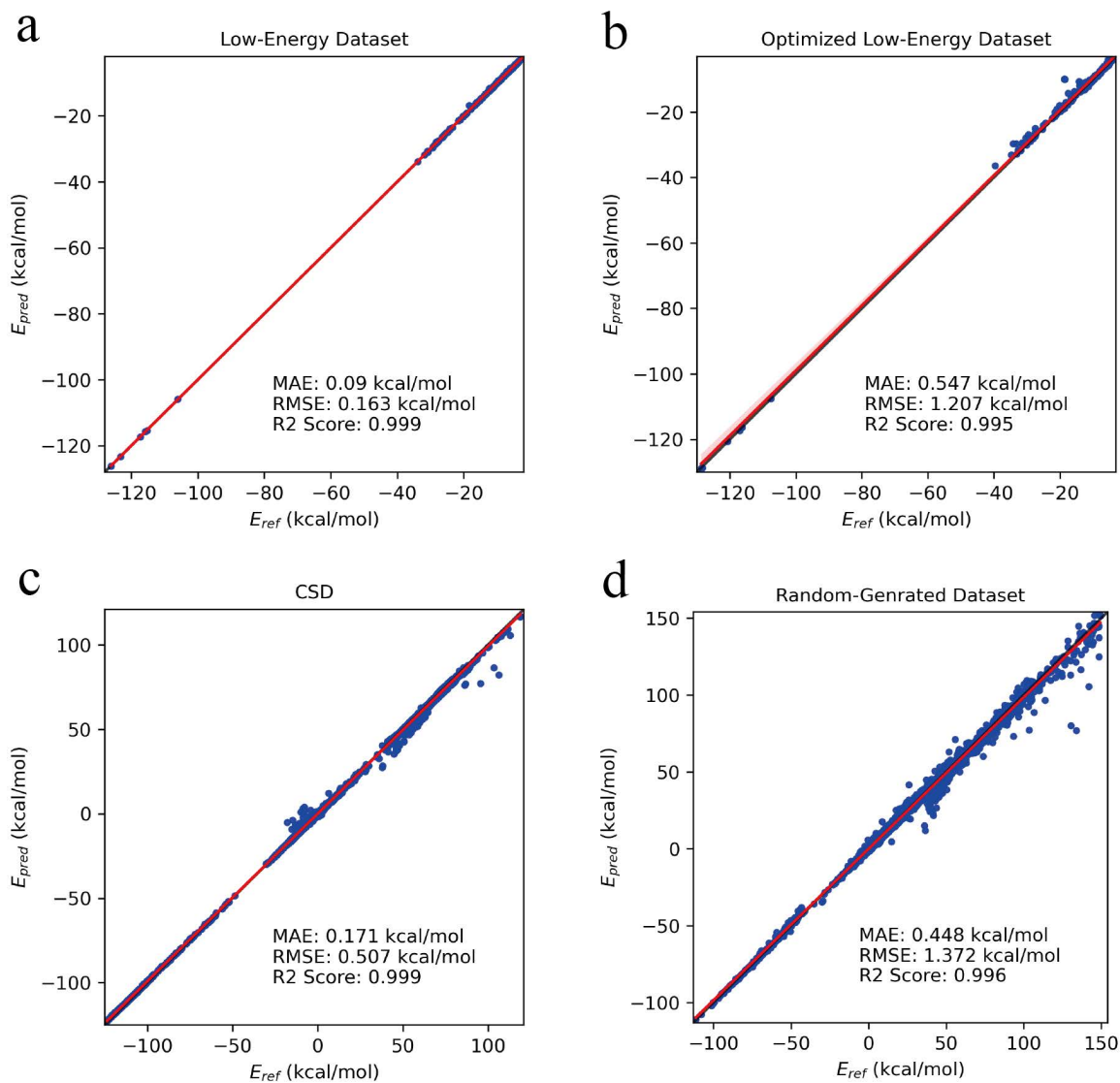

**Figure 2.** Correlation plots between non-covalent interaction energies calculated at ωB97X-D3BJ/def2-TZVPP level, $E_{ref}$, and predicted by PANIP, $E_{pred}$, for the low-energy dataset (a), optimized low-energy dataset (b), CSD subset (c), and random-generated dataset (d). The black line indicates perfect prediction (y=x), and the red line shows the regression fit.

**Comparison with ANI-2x.** Benchmarking against ANI-2x on the CSD-derived geometries (Figure 3a, b) and the other three benchmark sets (Figure S5) highlights PANIP's superior performance. On the full CSD subset, ANI-2x exhibited large errors (MAE 9.261 kcal/mol, RMSE 23.765 kcal/mol, $R^2$=0.064), with systematic overestimation of strongly attractive dimers (interaction energies < -40 kcal/mol) and underestimation of strongly repulsive dimers (> 40 kcal/mol), particularly in the presence of charged species (e.g., ACET–ETAM, ACET–ACET). These deviations likely stem from the absence of charged species in the ANI-2x's training data and the indirect computation of interaction energies as differences of total energies without explicit BSSE correction[36]. Even after excluding dimers containing charged fragments, ANI-2x remained less accurate (MAE 0.882 kcal/mol, RMSE 1.25 kcal/mol, $R^2$ = 0.846; Figure 3b), and similar performance limitations were observed across additional benchmark sets, where MAEs typically exceeded 1 kcal/mol (Figure S5). In contrast, PANIP directly predicts high-level NCI energies, avoids compounded errors from total energy differences, and consistently achieves sub-chemical-accuracy MAEs, including for charged and strongly interacting dimers. These results highlight the importance of high-fidelity NCI training data and a protein-specific design for accurate modeling of biomolecular NCIs.

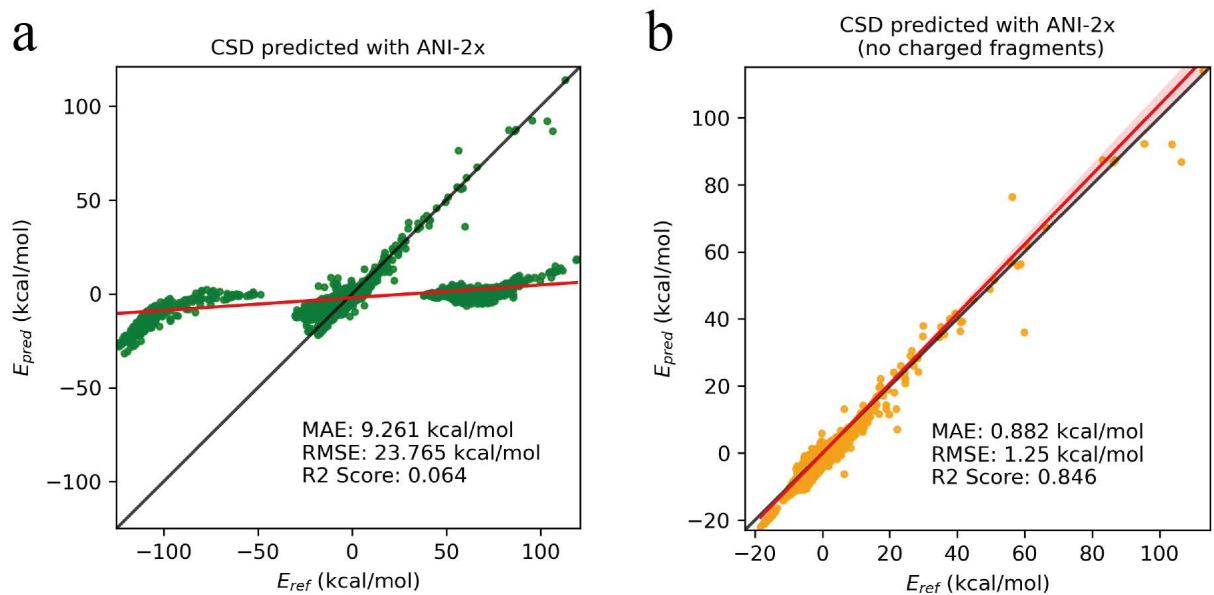

**Figure 3.** Correlation plots between non-covalent interaction energies calculated at ωB97X-D3BJ/def2-TZVPP level, $E_{ref}$, and predicted by ANI-2x, $E_{pred}$, for the CSD-derived benchmark set: (a) all samples; (b) excluding dimers containing charged fragments. The black line indicates perfect prediction (y=x), and the red line shows the regression fit.

**Exploring NCI Patterns in the PDB.** Leveraging PANIP's accuracy and efficiency, the interaction energies were predicted for all 36.3 million PDB-derived dimers, enabling systematic mapping of energy-geometry correlations across the entire dataset (Figure 4, Figure S6). Representative analyses focus on both well-characterized interactions like cation-π interactions, and previously underexplored interactions like dimethyl sulfide–aromatic contacts, to illustrate the model's ability to recover known patterns and reveal new trends. The lowest-energy representative conformations were selected for each interaction pattern and their NCI maps[37] were visualized to highlight regions where weak interactions predominantly occur, a detailed description of the NCI visualization method is provided in the Supporting Information.

For Cation-π Interaction, dimers involving lysine (ETAM) with tyrosine (PMPO) or tryptophan (MIND) were examined, yielding 28,914 unique ETAM-PMPO and 24,098 ETAM-MIND dimers. Spatial distribution of the cation relative to the aromatic ring revealed distinct low-energy geometries with centroid distances varying from 4.5 to 7 Å. Positively charged ETAM monomers are predominantly localized around the phenolic hydroxyl group of PMPO, forming a ring-shaped low-energy region, as well as another low-energy region above the aromatic π ring (Figure 4a and Figure S6a). Three representative conformations were identified. *Struc_1* with the lowest-energy (-20.41 kcal/mol) features ETAM simultaneously engaging in a cation-π interaction (N atom 2.9 Å from the phenol ring center) and a nonclassical hydrogen bond between the methyl group of ETAM and the hydroxyl group of PMPO. Cation-π interaction dimers accounted for 17.7% of ETAM-PMPO dimers. *Struc_2* (-19.52 kcal/mol) is dominated by a strong NH⋯O hydrogen bond between ETAM's NH group and PMPO's hydroxyl oxygen, with a prevalence of 19.8%. The conformation involving a CH⋯π interaction between the methylene groups of ETAM and the aromatic ring of PMPO is less frequent and higher in energy (Figures 4a, *struc_3*).

Due to the enhanced electron density on Trp's six-membered ring, ETAM tends to interact with MIND above this ring (Figure S6b), with centroid distances of 5-6.5 Å. Figure 4b shows the lowest-energy structures of three representative NCI patterns. The lowest-energy structure, *struc_1* (-24.22 kcal/mol), involves a strong cation-π interaction between ETAM's nitrogen atom and the indole ring. Cation-π interaction dimers accounted for 27.6% of ETAM-MIND dimers. Conformation represented by *struc_2* (-15.44 kcal/mol) adopts a "parallel" orientation between ETAM heavy-atom plane and the indole ring, in which hydrogens attached to the three heavy atoms form weak XH⋯π interactions with the aromatic ring simultaneously. In contrast, *struc_3* (-10.83 kcal/mol) features a perpendicular arrangement dominated by methylene CH···π interactions accounting for 33.7% of dimers. These findings align with prior studies[38–42], supporting the reliability of PANIP for large-scale NCI analysis.

For previously underexplored dimethyl sulfide–aromatic Interactions[43,44], 78,674 MSM (methionine)–MBZ (phenylalanine) dimers were extracted and analyzed. These revealed five distinct low-energy geometric patterns (Figure 4c, Figure S6c, and Figure S7), spanning interaction distances of 3.5–7 Å with a predominant peak near 5.1 Å and interaction energies ranging from -4.9 to 0.2 kcal/mol (most populated around -1.6 kcal/mol). Mapping the sulfur positions shows that low-energy structures (~ -4 kcal/mol) concentrate in a previously unreported ring-shaped region above the aromatic plane, reflecting stereochemical preferences of sulfur and complementary electrostatic interactions (Figure S6c). Joint analysis of energetic and orientational preferences identifies five low-energy patterns. In the innermost region, C–S bond points toward the ring (Figure S7a), consistent with stereochemical preferences of divalent sulfur approaching the positively polarized aromatic edge (Figure 4c, *struc_5*, -1.52 kcal/mol). Parallel arrangements, in which both the C–S bond and the MSM plane align parallel to the aromatic ring (Figure S7b and S7c), yield the lowest energy (*struc_1*, -4.91 kcal/mol) through complementary electrostatic interactions between sulfur and methyl dipoles. When positioned farther from the aromatic ring, MSM assumes an outward-pointing C–S bond geometry (Figure S7a; *struc_4*, -3.06 kcal/mol), consistent with π electrons interacting with the C–S σ antibonding orbital. A fourth pattern places MSM above the ring with a perpendicular orientation, where one C–S bond aligns antiparallel to the ring dipole (*struc_2*, -4.43 kcal/mol). Geometries featuring C–H···S contacts are also observed (*struc_3*, -3.21 kcal/mol), reflecting weak hydrogen-bond donation from aromatic C–H groups to sulfur lone pairs. Collectively, these motifs underscore how methionine-like sulfur can engage aromatics through directional stereoelectronic and electrostatic complementarity, providing structural principles relevant to stereospecific recognition in Met-containing binding environments.

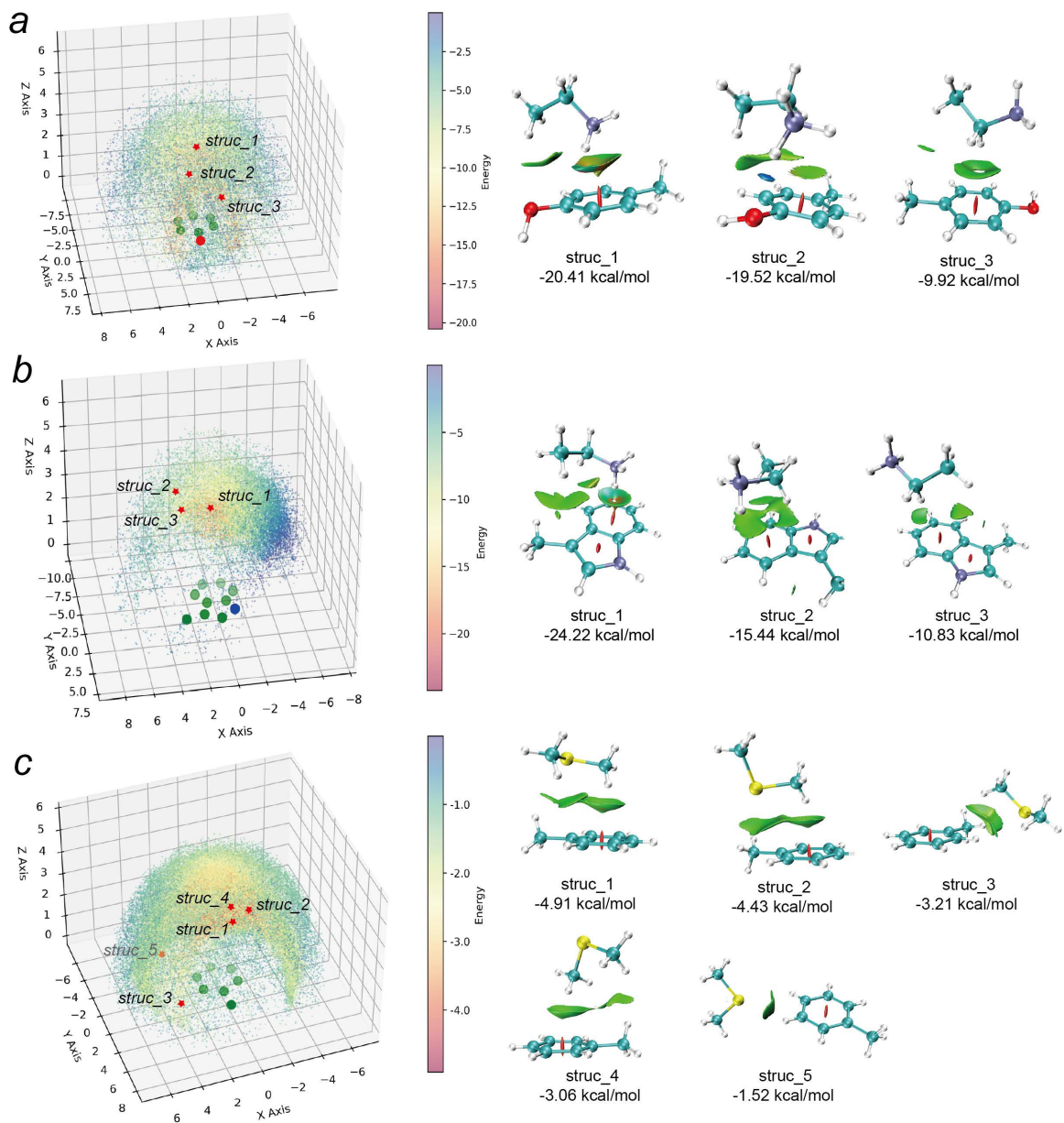

**Figure 4**. Spatial distribution and representative low-energy structure of ETAM-PMPO (a), ETAM-MIND (b), and MBZ-MSM (c) dimers. In the left column, each scatter point represents the position of the fragment's central atom (N or S) relative to the central fragment. The color of each scatter point corresponds to the energy scale shown in the adjacent color bar. Circles on the xy-plane represent heavy atoms of the central fragment: green for carbon atoms, red for oxygen atoms, and blue for nitrogen. Red stars mark the positions of the structures shown in the right column. The right column depicts the lowest-energy structures representing characteristic interaction patterns for each dimer. Isosurfaces show NCI maps between the fragments, and the NCI energy of each structure is given below the corresponding image.

**Application to Protein-Ligand Complexes.** To demonstrate the potential application in biomolecular modeling, PANIP was applied as a fragment-based scoring function for protein-ligand docking on three model systems: (1) indole bound to the apolar L99A mutant (PDB: 185L)[45], (2) phenol bound to the polar L99A/M102H double mutant (PDB: 4I7L)[46], and (3) serine bound to pyruvate kinase M2 (PKM2, PDB: 4B2D)[47] (Figure 5). The interaction energies $\Delta E_{bind}$ of all docking poses were calculated using both our PANIP scoring function (Eq.4) and the united AMBER force field energy function in DOCK. Pose prediction accuracy was quantified via root mean square deviation (RMSD) of ligand heavy atoms relative to crystallographic poses (Table 1). Native poses extracted from crystal structures were included in the docking pool for validation.

Across these systems, PANIP substantially improved pose ranking relative to an AMBER-based docking score. In L99A-indole, PANIP ranked the native pose first among 500 docking poses, while DOCK score selected a pose with an RMSD of 0.34 Å (ranked No. 160; Figure 5a, b). In L99A/M102H-phenol, PANIP identified the native pose (No. 1), outperforming DOCK (No. 76; RMSD 0.21 Å; Figure 5c, d). In PKM2-serine, the serine ligand was decomposed into fragments (ACET, ETAM, and ETOH) for binding energy estimation. PANIP ranked the native pose third (RMSD 0.44 Å), while DOCK ranked it No. 46 (RMSD 1.29 Å). The top PANIP pose features a hydrogen bond (2.321 Å) between ligand serine's hydroxyl group and the carbonyl group of ILE469, differing from the native pose (2.149 Å; OH → ARG43 C=O; Figure 5e). However, the top PANIP-selected pose positioned the serine hydroxyl group near the serine NH3+ fragment, potentially inducing intramolecular repulsion not captured by the pairwise fragment approximation, illustrating a key limitation of the current framework.

**Table 1.** Binding pose prediction for three protein targets: L99A (PDB 185L) in complex with indole; L99A/M102H in complex with phenol (PDB 1LI2); PKM2 in complex with serine (PDB 4B2D) with PANIP and AMBER. The native pose is the pose in the crystal structure. The best prediction is the best pose predicted by different scoring functions.

| Ligand | Protein | PDB ID | Ranking of Native Pose (500 poses) | | RMSD of the best prediction (Å) | |
|---|---|---|---|---|---|---|
| | | | PANIP | DOCK | PANIP | DOCK |
| Indole | L99A | 185L | 1 | 160 | 0.10 | 0.34 |
| Phenol | L99A/M102H | 4I7L | 1 | 76 | 0.18 | 0.21 |
| Serine | PKM2 | 4B2D | 3 | 46 | 0.44 | 1.29 |

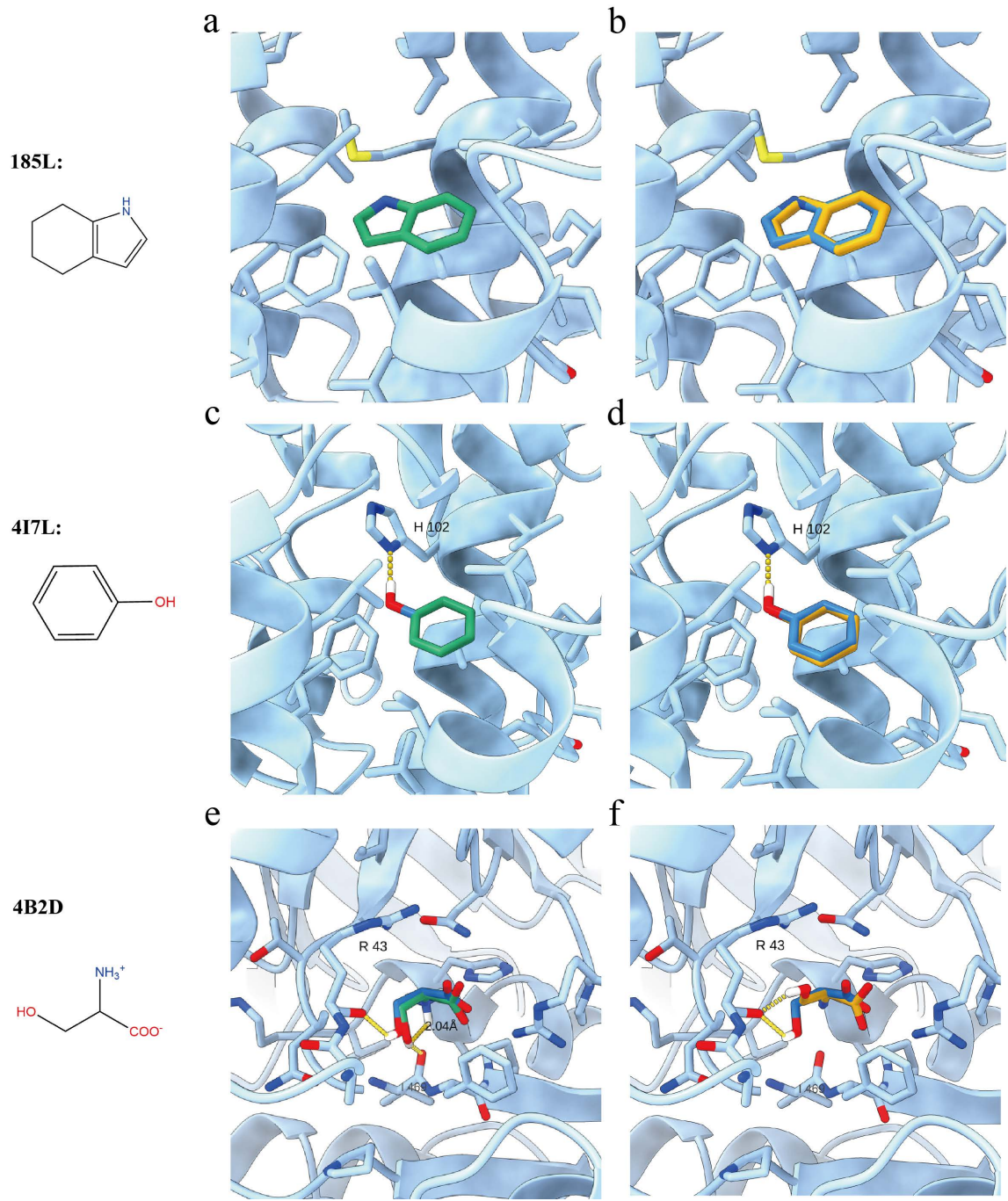

**Figure 5.** Structures and the best binding pose of indole ligand in L99A (top), phenol ligand in L99A/M102H (middle), and serine in PKM2 (bottom) predicted by our scoring function (green) and AMBER (yellow). The native structures are colored in blue for comparison. Structures are visualized in ChimeraX[76].

Additional model systems were evaluated (Table 2), including various alkyl-substituted benzene isomers and phenol binding to L99A and its polar or charged variants (L99A/M102Q, M102H, and M102E), as well as amino acid ligands such as alanine and phenylalanine interacting with PKM2. Across these systems, PANIP-based scoring function achieved a 44% native pose recovery rate (ranked first) across 25 model systems, consistently outperformed the DOCK score in recovering native poses and reducing RMSDs of top-ranked structures. Notably, this performance is achieved without explicit long-range electrostatic or solvation corrections, suggesting that accurate short-range NCI energetics are a dominant factor for pose discrimination in these relatively simple model systems.

**Table 2.** Binding pose prediction for three protein targets: L99, L99A/M102Q, and PKM2 with PANIP and DOCK. The native pose is the pose in the crystal structure. The best prediction is the best pose predicted by different scoring functions*.

| Ligand | PDB ID | Ranking of Native Pose (500 poses) | | RMSD of the best prediction (Å) | |
| --- | --- | --- | --- | --- | --- |
| | | PANIP | DOCK | PANIP | DOCK |
| L99A | | | | | |
| toluene | 4W53 | 82 | 128 | 0.29 | 0.41 |
| isobutylbenzene | 184L | 1 | 271 | 0.37 | 0.91 |
| n-Butylbenzene | 4W57 | 1 | 34 | 1.36 | 1.38 |
| benzene | 181L | 95 | 490 | 0.93 | 0.83 |
| benzene | 3HH4 | 40 | 250 | 0.11 | 0.64 |
| benzene | 4W52 | 57 | 273 | 0.06 | 0.67 |
| ethylbenzene | 4W54 | 1 | 500+ | 1.47 | 1.57 |
| ethylbenzene | 3HH6 | 2 | 250 | 0.28 | 0.50 |
| propylbenzene | 4W55 | 33 | 500+ | 0.26 | 0.39 |
| sec-butylbenzene | 4W56 | 1 | 11 | 0.99 | 1.07 |
| octylbenzene | 4W59 | 58 | 231 | 2.004 | 2.01 |
| p-toluene | 187L | 95 | 470 | 0.55 | 0.58 |
| L99A/M102Q | | | | | |
| benzene | 5JWT | 71 | 197 | 0.23 | 0.59 |
| ethylbenzene | 5JWV | 1 | 198 | 0.68 | 0.91 |
| phenol | 1LI2 | 64 | 69 | 0.27 | 0.15 |
| L99A/M102H | | | | | |
| benzene | 4I7J | 33 | 323 | 0.48 | 0.50 |
| toluene | 4I7K | 37 | 284 | 0.33 | 0.64 |
| L99A/M102E | | | | | |
| benzene | 3GUJ | 1 | 269 | 0.20 | 0.39 |
| toluene | 3GUK | 1 | 223 | 1.46 | 0.65 |
| ethylbenzene | 3GUL | 6 | 95 | 0.22 | 0.36 |
| PKM2 | | | | | |
| alanine | 2G50 | 9 | 30 | 0.71 | 0.71 |
| phenylalanine | 4FXJ | 1 | 1 | 0.43 | 0.45 |

\* For structures exhibiting alternative conformations, the scoring results correspond to the conformation with the highest occupancy value.

DISCUSSION

We present PANIP, a protein-specific MLIP for NCIs, trained on PDB-derived fragment dimers with ωB97X-D3BJ/def2-TZVPP reference energies. A multi-fidelity active learning workflow enabled the construction of PDB-FRAGID, a high-precision, low-redundancy dataset that compresses 36.3 million dimers to ~ 3.15 million representative structures while preserving the diversity of 17 fragment types and 153 dimer combinations. PANIP achieves sub-chemical-accuracy MAEs across multiple benchmark sets, generalizes to CSD-derived and non-equilibrium geometries, and offers near force-field computational cost, making it suitable for large-scale exploration of protein NCI landscapes.

Compared to the widely used ANI-2x potential, PANIP delivers markedly lower errors on protein-derived fragments, especially for charged and strongly interacting dimers, underscoring the value of targeted, high-fidelity NCI training data. When integrated into a fragment-based energy decomposition scheme, PANIP serves as an effective scoring function for protein–ligand docking, outperforming an AMBER-based score in pose ranking for a range of model systems. While the current framework focuses on pairwise interactions, ongoing efforts aim to expand fragment diversity to cover broader chemical spaces, to incorporate long-range electrostatic, multi-body, and solvation effects. Strategies include integrating well-established force field correction terms[48,49], leveraging ML-based correction schemes[50,51], or employing metamodeling approaches[52]. Together, PDB-FRAGID and PANIP provide a foundation for scalable, QM-accurate modeling of protein NCIs and for refining both data-driven and classical force-field descriptions in biomolecular simulations.

METHODS

**Overall Workflow**. As illustrated in Figure 6, a systematic workflow was developed to build MLIPs that leverage chemically relevant fragment-fragment interactions derived from high-resolution protein structures in the PDB. Proteins were fragmented into dimeric pairs to generate an initial dataset, which was first labeled with baseline interaction energies computed at the low-cost r2SCAN-3c level[28]. An MFAL method was designed to iteratively guide the selection of a representative and diverse subset of dimers, ensuring balanced coverage of the full structural distribution. This refined subset, termed the PDB-FRAGID, was recalculated at the higher-accuracy ωB97X-D3BJ/def2-TZVPP level[29,30]. The curated data was then used to train the PANIP models, resulting in robust MLIPs with QM level accuracy in predicting molecular interaction energies.

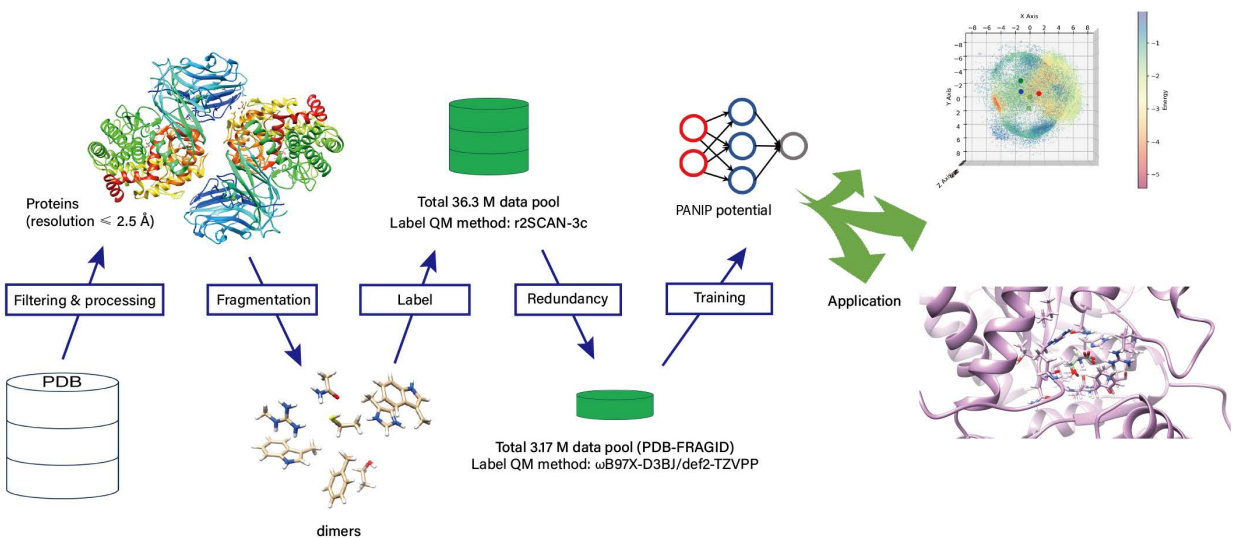

**Figure 6.** Overall workflow of dataset construction and model training.

**Fragmentation Definition and Dimer Extraction.** To represent NCIs in proteins, 17 chemically distinct fragment types were defined, encompassing amino acid side chains, backbone segment, and water (Figure 7). Key atoms (e.g., carbons bonded to aromatic rings, thiols, hydroxyls, or ammonium groups) were retained to preserve local electronic environments and minimize symmetry-related redundancy. These fragments were paired into 153 unique dimer types, forming the basis for NCI analysis.

Initial protein structures were obtained from PDB entries with resolution ≤ 2.5 Å and unique UniProt[53] IDs to reduce redundancy, yielding 29,204 proteins. Missing hydrogens were added using the high-throughput molecular dynamics (HTMD) approach[54], while the original PDB coordinates were preserved without further structural optimization. Protonation states of ionizable residues were assigned according to typical physiological conditions: all three common protonation states were considered for histidine, reflecting its pKa near physiological pH, whereas lysine and arginine side chains were treated as protonated, and aspartic acid side chains as deprotonated.

Interacting fragment pairs were identified using two criteria. First, the shortest heavy-atom distances between 2 and 4 Å (capturing interactions slightly beyond the sum of van der Waals radii for atomic pairs (C=1.70 Å, N=1.55 Å, O=1.52 Å, S=1.80 Å) and excluding covalent bonds (<2 Å)[55]). The 4 Å upper bound was chosen to focus on geometries where NCIs are most significant, while avoiding an overwhelming number of weak, long-range contacts. Second, fragment pairs were required to be separated by at least two residues along the protein sequence to mitigate trivial local contacts arising from backbone connectivity. This yielded a raw set of 36.3 million dimers spanning 153 unique fragment-fragment combinations across the 17 fragment types in protein (see Supporting Information for details).

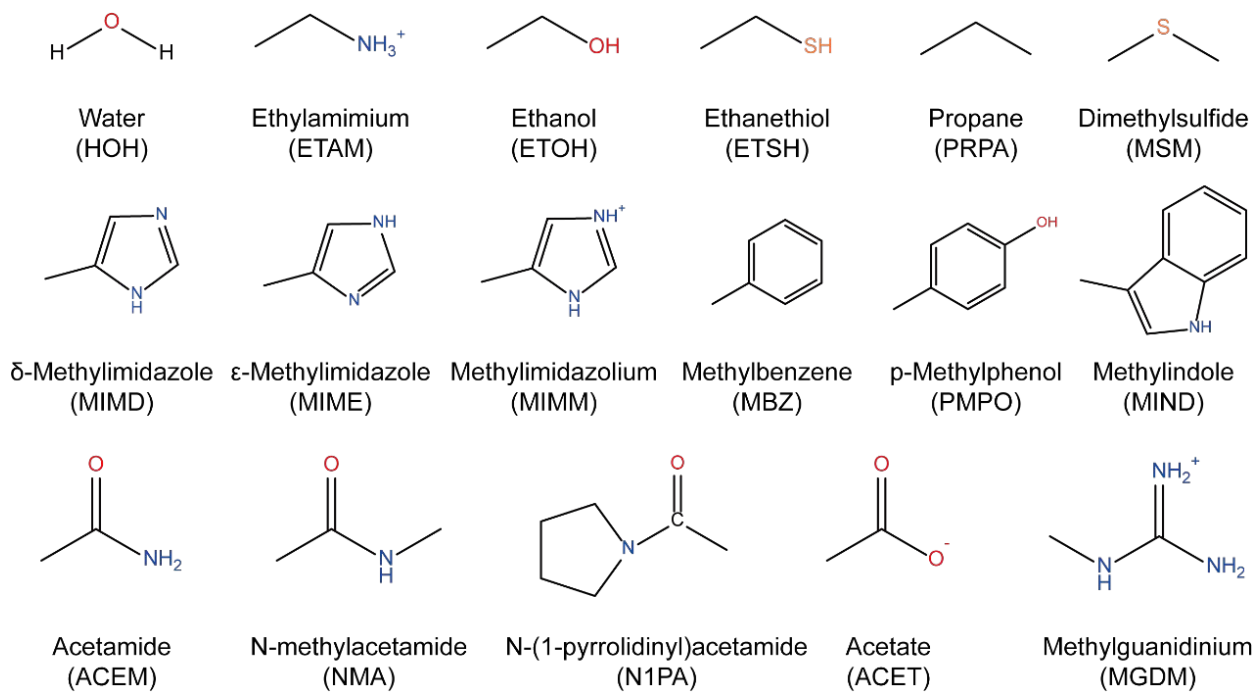

**Figure 7.** 17 representative chemical groups defined in proteins.

**Data Reduction via Multi-Fidelity Active Learning.** To balance computational efficiency with accuracy, a multi-fidelity active learning workflow[26,27] was designed (Figure S1). MFAL extends active learning by combining low-cost quantum chemical approximations with selected high-fidelity refinements, reducing the need for exhaustive expensive calculations. Specifically, the r²SCAN-3c method[28] served as a low-fidelity oracle to provide baseline interaction energies for all dimers. r²SCAN-3c combines the r²SCAN meta-GGA functional with atom-pairwise dispersion and basis-set corrections, and was calculated using ORCA 5.0.3[56]. The non-covalent interaction energy ($\Delta E_{AB}$) for each dimer was computed as:

$$\Delta E_{AB} = E(AB) - [E(A) + E(B)] \qquad (1)$$

where $E(AB)$, $E(A)$, and $E(B)$ represent the total energy of the dimer and isolated monomers, respectively.

As an initial step, a 2% random bootstrap set was used to train an initial surrogate model based on NequIP[31]. The surrogate model was then iteratively applied to the remaining dimers, and an uncertainty-based acquisition function flagged high-error cases using the normalized criterion $|E_{pred} - E_{ref}|/\sqrt{N} > 0.04$ kcal/mol, where $E_{pred}$ is the predicted energy, $E_{ref}$ is the reference (r²SCAN-3c) energy, and $N$ is the atom count. In each iteration, a random 2% subset of these flagged structures was added to the training set. This iterative loop continued until fewer than 5% of the unscreened dimers remained as high-error cases, which were then fully included in the final dataset. The final refined subset (PDB-FRAGID) condensed the original pool to a representative core comprising only 8.7% of its size (~3.15 million) while preserving balanced coverage of the full structural distribution.

**High-Level QM labeling.** All dimers in the PDB-FRAGID dataset were recomputed at the higher-accuracy ωB97X-D3BJ/def2-TZVPP level[29,30] using ORCA 5.0.3[56]. The ωB97X-D3BJ[29] (a range-separated hybrid meta-GGA density functional with D3(BJ) empirical dispersion correction) coupled with the def2-TZVPP[30] basis set delivers exceptional precision for modeling non-covalent interactions and conformational energies while maintaining a balanced treatment of short- and long-range electronic correlations[29,30]. The resolution of identity approximation with coulomb, exchange, and correlation contributions (RIJCOSX) approximation[57] was applied to accelerate calculations, with auxiliary basis sets def2/J[58] and def2-TZVPP/C[59] employed for integration and correlation treatments. The ma-def2-TZVPP[60] diffuse function was added only for negatively charged oxygen atoms (e.g., in carboxylate groups like ACET) to improve electron density descriptions. For neutral systems, additional diffuse functions provided no benefit and could even introduce numerical instabilities. The basis set superposition error (BSSE) was corrected using the counterpoise method[61–63]. NCI energies were computed according to Eq.1. These high-fidelity ωB97X-D3BJ/def2-TZVPP interaction energies on PDB-FRAGID serve as the reference data for training PANIP.

**Training Protocol of PANIP.** PANIP was implemented as an ensemble of NequIP[31], an equivariant graph neural network that explicitly encodes the rotational, translational, and permutational symmetries of atomistic systems. By enforcing E(3)-equivariance in its message-passing layers, NequIP achieves high data efficiency and systematically improved accuracy compared to conventional invariant graph neural networks. This makes it particularly well-suited for learning NCI energies, where capturing subtle orientation-dependent non-covalent effects is essential. For NequIP training, we modified the default settings as follows: a cutoff radius of 7.5 Å, inclusion of O, N, and S in the chemical symbols list, and a batch size of 512. Early stopping was applied with a patience of 100 epochs on the validation loss. Global energy rescaling was set to *dataset_total_energy_std* (the standard deviation of total energies in the dataset).

To accelerate the construction of the final model, MFAL-selected subsets from each fragment type were first used to train fragment-specific NequIP models. These fragment-wise datasets were then combined to form the unified PDB-FRAGID dataset, which was used to train the final PANIP ensemble (5 models) via five-fold cross-validation. Unless otherwise noted, predictions are reported as ensemble averages. In addition, fragment-specific models generally exhibit slightly higher accuracy on their respective subsets but reduced generalization across fragment types, whereas the unified PANIP model provides robust performance across the full diversity of protein-derived NCIs. Full training protocols, hyperparameters, and model files are available on GitHub (https://github.com/hnlab/PANIP).

**Benchmark Datasets.** PANIP was evaluated on four benchmark sets derived from both protein and non-protein sources. First, the low-energy representative from the original r2SCAN-3c-labeled pool was selected for each unique dimer type, resulting in 277 candidate structures. These structures were optimized at the r²SCAN-3c level to ensure well-defined equilibrium geometries, forming a dedicated benchmark set. Single-point energy calculations were performed using the ωB97X-D3BJ/def2-TZVPP method on all selected dimer geometries (with and without optimization). This benchmark set was used to assess model performance in reproducing equilibrium interaction energies. Same-charge dimer pairs were excluded due to frequent wave function convergence failures during optimization.

Second, to evaluate model transferability beyond training data, dimers involving the same fragment types were extracted from CSD[32] using ConQuest[64]. Although the chemical fragment pairs are identical, the geometries originate from small-molecule structures (not protein environments) and thus sample distinct conformational and packing motifs—providing a complementary test of generalization across structural sources. To preserve structural integrity and chemical independence, only dimers featuring single aliphatic bonds (C-C or C-H) were retained. Pairs with the closest interatomic distances between 2 and 4 Å were selected, aligning with NCI criteria established for PDB dimers. Organometallic structures were excluded and hydrogen positions were normalized to standard bond geometries, yielding 33,274 CSD-derived dimers for analysis.

Third, to evaluate performance on non-equilibrium conformations, a biased fragment-pair random sampling method was developed. Fragments were treated as rigid bodies with fixed bond lengths and angles. Initial 3D fragment structures were generated from SMILES strings[65] using the RDKit software package[66]. One fragment was fixed at the origin, and the other was positioned within a 12 Å radius sphere centered on the fixed fragment, with rotational sampling applied. A biased random sampling protocol was implemented using a logarithmic normal distribution for the acceptance probability:

$$x = r_i^{vdw} + r_j^{vdw} - d_{ij} \qquad (2)$$

$$p(x) = \frac{1}{(-x+c)\sigma\sqrt{2\pi}} e^{-\frac{(\ln(-x+c)-\mu)^2}{2\sigma^2}} \qquad (3)$$

Where $x$ is the intermolecular overlap (Å), $r_i^{vdw}$ and $r_j^{vdw}$ are the van der Waals radii of atoms $i$ and $j$, and $d_{ij}$ is their interatomic distance. Parameters were set to σ=0.8, μ=1.0, and c=1.6 —chosen to maximize acceptance probability at an intermolecular overlap of ~0.1 Å (favoring physically realistic conformations) and ensure overlaps remained below 1.6 Å to avoid severe clashes[67,68] (Figure S2). This coordinate system enabled denser sampling at shorter intermolecular distances and sparser sampling at longer distances. For each dimer type, 100 conformations were sampled, yielding a total of 15,300 conformations. Interaction energies for these dimers were computed using both PANIP and ωB97X-D3BJ/def2-TZVPP for comparison.

Finally, to benchmark against an established ML potential, the ANI-2x neural network potential[9] (a widely used model trained on large organic molecule datasets, covering C, H, O, N, and S elements) was employed to predict NCI energies of the CSD subset for comparison with PANIP. The pre-trained ANI-2x neural networks are available in TorchANI python library[69]. Since ANI-2x predicts total energies, NCI energies were calculated by applying Eq. 1 to the ANI-2x predicted total energies of the dimer and its corresponding monomers. Because ANI-2x was not parameterized for charged systems, performance was evaluated both on datasets including and excluding charged dimers, with detailed results provided in the Supporting Information.

**Docking and Rescoring.** We evaluated the model ensemble's ability to predict the native-like binding pose for three protein-ligand model systems[70,71]: L99A in complex with indole (PDB 185L)[45], L99A/M102H in complex with phenol (PDB 4I7L)[46], and the M2 isoform of pyruvate kinase (PKM2) in complex with serine (PDB 4B2D)[72]. Docking was performed using default parameter settings from an automated platform as described previously[73,74]. For all docking poses generated by DOCK 3.7[75], binding energies were first calculated using scoring functions based on the united AMBER force field (as in DOCK 3.7).

To compute binding interaction energies with PANIP, both the receptor and ligand were fragmented within a 5 Å distance cutoff around the ligand (Figure 8). Notably, for PKM2-serine, the serine ligand was decomposed into three fragments (ACET, ETAM, and ETOH). The PANIP-estimated binding interaction energy is given by

$$\Delta E_{bind} = \sum_{i}^{rec} \sum_{j}^{lig} \Delta E_{pred} + \Delta E_{lnk} - \Delta E_{dup} \qquad (4)$$

where $\Delta E_{pred}$ is the interaction energy between protein fragment $i$ and ligand fragment $j$ (computed using PANIP), $\Delta E_{lnk}$ accounts for energy contributions from linking atoms (protein/ligand, green circle in Figure 8), and $\Delta E_{dup}$ corrects for double-counted atoms in fragment-fragment interactions (orange circle in Figure 8).

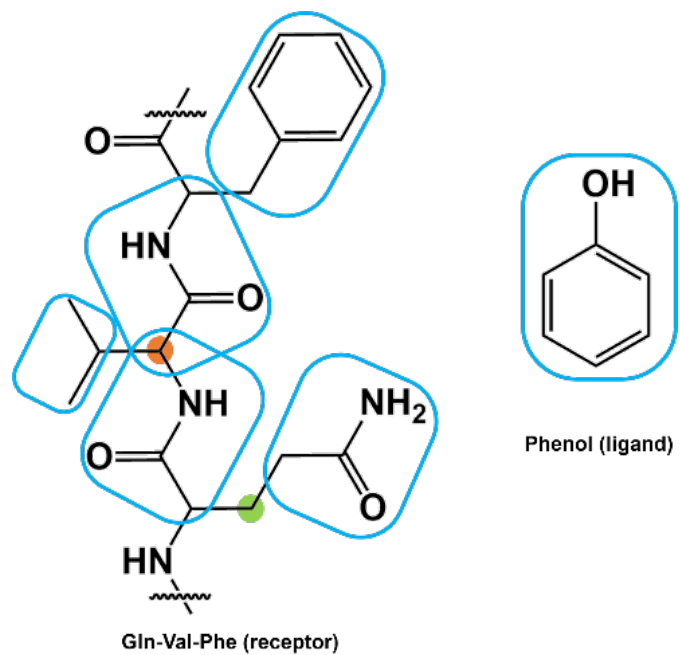

**Figure 8.** The sketch of protein fragmentation approach. Receptor and ligand are decomposed into fragments (blue boxes), linker atoms (green circle), and duplicated atoms (orange circle)

AUTHOR INFORMATION

**Corresponding Author**

*E-mail: huangniu@nibs.ac.cn

**ORCID**

Niu Huang: 0000-0002-6912-033X

Lejia Zeng: 0009-0004-3410-8404

**Funding Sources**

Beijing Municipal Science and Technology Commission, Administrative Commission of Zhongguancun Science Park, Grant/Award Number: Z201100005320012

**Notes**

The authors declare no competing financial interest.

ACKNOWLEDGMENT

This work is supported by Beijing Municipal Science & Technology Commission (Z201100005320012 to Niu Huang) and Tsinghua University.

ABBREVIATIONS

AL, active learning; BSSE, basis set superposition; CSD, Cambridge Structural Database; HTMD, high-throughput molecular dynamics; MAE, mean absolute error; MLIPS, machine learning interatomic potentials; ML, machine learning; NCIs, non-covalent interactions; NequIP, Neural Equivariant Interatomic Potentials; PDB, Protein Data Bank; PDB-FRAGID, PDB Fragment Interaction Dataset; PES, potential energy surfaces; PKM2, M2 isoform of pyruvate kinase; RMSD, root mean square deviation; QM, quantum mechanical.

TOC Graphic

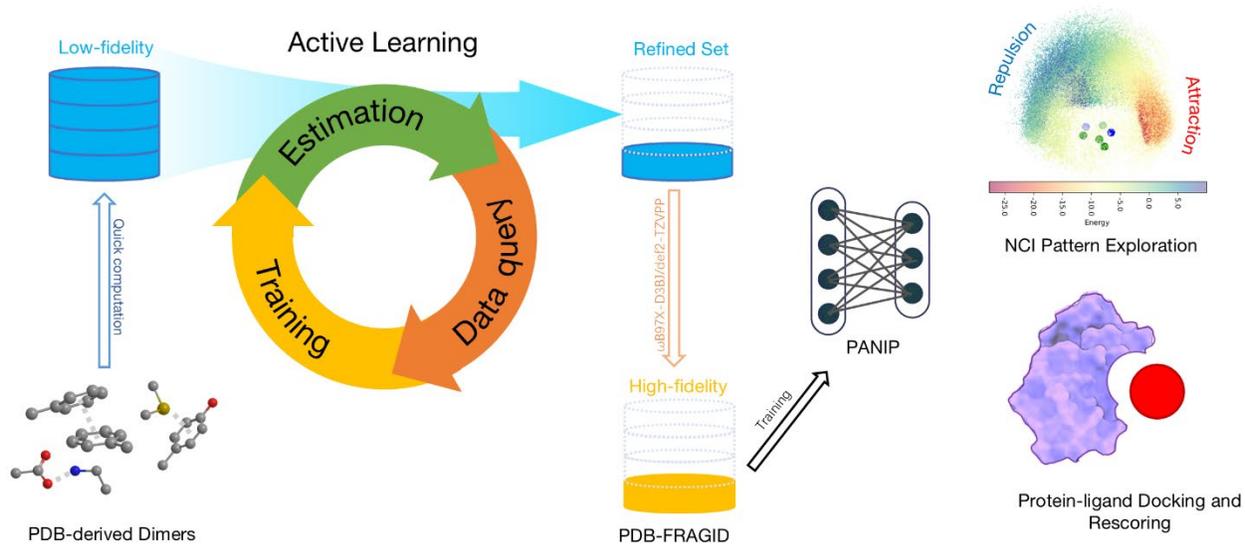

Robust MLIPs that achieve accuracy comparable to the ωB97X-D3BJ/def2-TZVPP level on non-covalent interactions.

# Supporting Information for Developing a Machine Learning Interatomic Potential for Non-Covalent Interactions in Proteins

*Lejia Zeng[1,2], Xintong Zhang[1,2], Yuchan Pei[1,2], Lifeng Zhao[2], Lan Hua[2], Jincai Yang[2], Niu Huang[1,2,*]*

[1]Tsinghua Institute of Multidisciplinary Biomedical Research, Tsinghua University, Beijing 102206, China

[2]National Institute of Biological Sciences, 7 Science Park Road, Zhongguancun Life Science Park, Beijing 102206, China

E-mail: huangniu@nibs.ac.cn

# Contents

# 1 Dimer Extraction

Dimers were extracted from protein structures using an in-house script. For each type of fragment, a corresponding SMARTS[1] pattern was generated, and SMILES[2] patterns were also employed to define the three-dimensional templates. Substructure searches across all collected proteins were then carried out through SMARTS-based substructure matching implemented in RDKit[3] Python package. The coordinates of matching atoms of each hit were mapped onto the template atoms defined by corresponding SMILES patterns. When isolation of the fragment required cleavage of a C–C or C–N bond, the bond was cut at the appropriate position, the resulting terminal atom was first marked as a dummy atom, and then capped with hydrogen to maintain valency.

Once 17 chemical groups had been collected from a given protein, dimers were searched for by identifying fragment pairs in close spatial proximity. Each interacting pair was initially labeled as the shortest distance between 2 and 4 Å. And the residue in which two fragments are located should be separated by at least 2 amino acids to avoid bias from proximity effects.

# 2 Active Learning Workflow

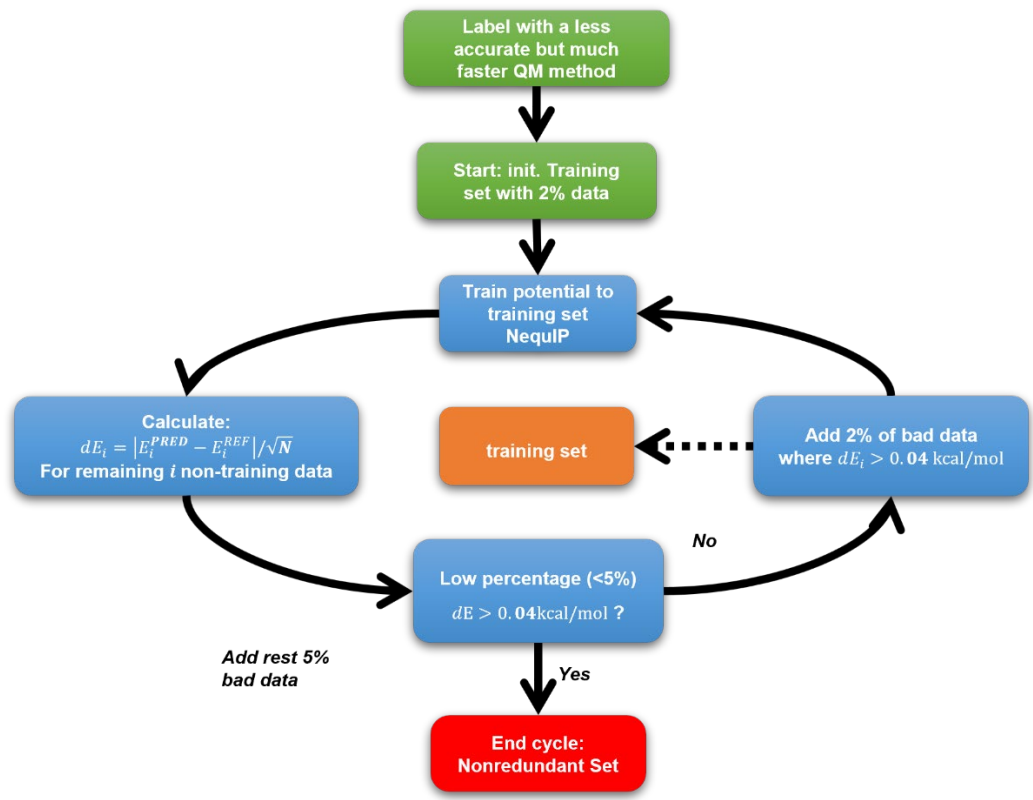

**Figure S1.** AL workflow for data redundancy.

## 3  Random Sampling

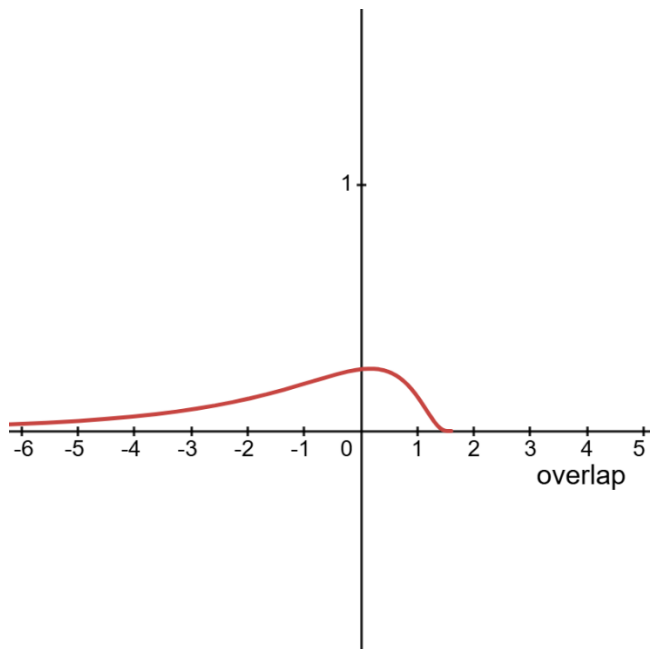

**Figure S2.** Acceptance probability curve for intermolecular overlap used in random sampling. Curve visualization generated via https://www.desmos.com/calculator/a8imk7uszg.

# 4 Model Performance on Benchmark Sets

a

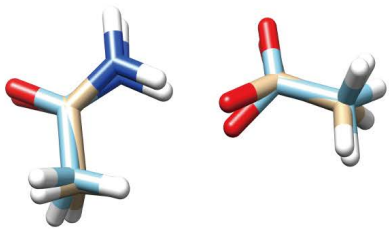

b

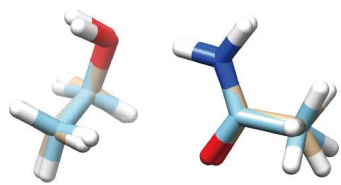

**Figure S3.** Conformational changes observed in post-optimization. Initial lowest-energy structures (tan) and their optimized counterparts (sky blue) exhibit minimal structural deviations, as quantified by root-mean-square deviation (RMSD) values of (a) 0.399 Å for ACEM-ACET and (b) 0.352 Å for ACEM-ETOH.

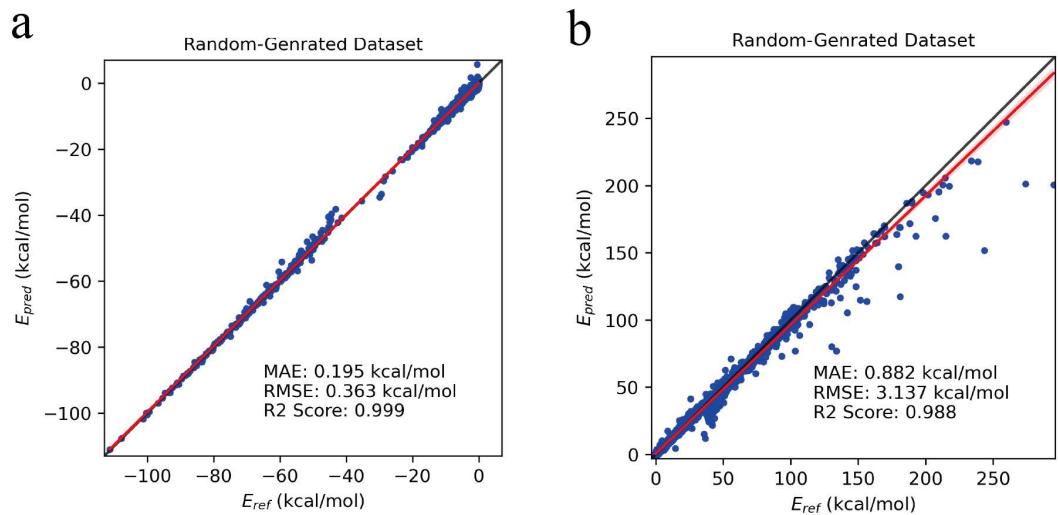

**Figure S4.** Interaction energies calculated at ωB97X-D3BJ/def2-TZVPP level and predicted by PANIP for samples with energy (a) below 0 kcal/mol and (b) above 0 kcal/mol in random-generated dataset. The black line indicates perfect prediction (y = x), and the red line shows the regression fit.

# 5 Comparison with ANI-2x

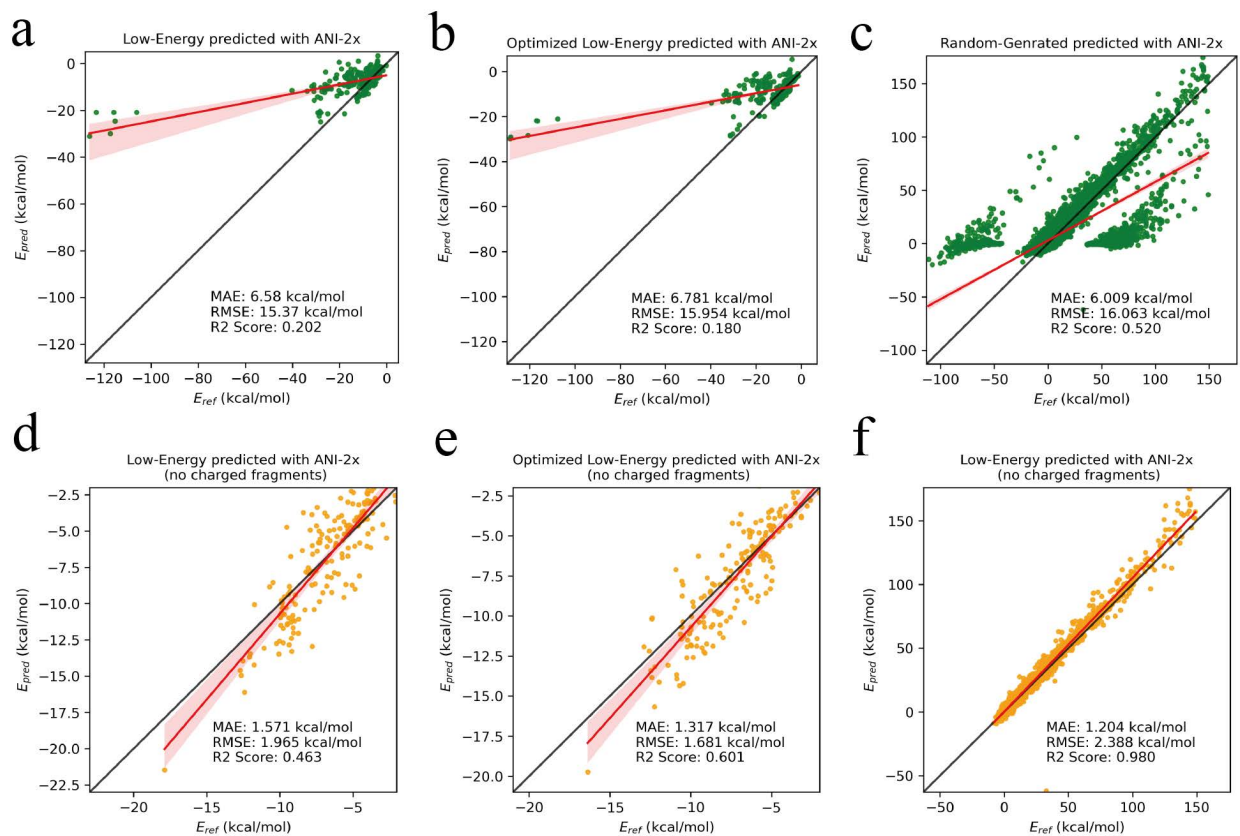

**Figure S5.** Correlation plots between non-covalent interaction energies calculated at ωB97X-D3BJ/def2-TZVPP level, $E_{ref}$, and predicted by ANI-2x, $E_{pred}$, for low-energy dataset (a, d), optimized low-energy dataset (b, e), and random-generated dataset (c, f). Panels a–c include all samples; panels d–f exclude dimers containing charged fragments.

# 6 Exploring NCI Patterns in PDB

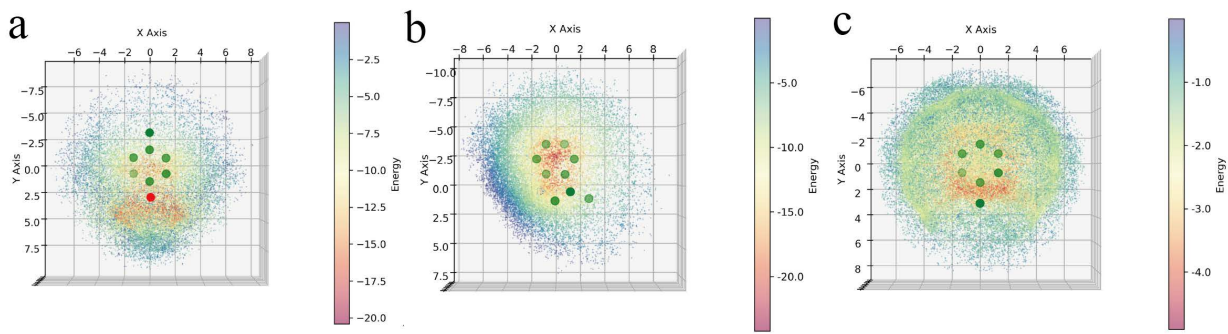

**Figure S6.** Bottom view of spatial distributions for ETAM-PMPO (a), ETAM-MIND (b), and MBZ-MSM (c). Each scatter point represents the position of the fragment's central atom (N or S) relative to the central fragment. The color of each scatter point corresponds to the energy scale shown in the adjacent color bar. Circles on the xy-plane represent heavy atoms of the central fragment: green for carbon atoms, red for oxygen atoms, and blue for nitrogen.

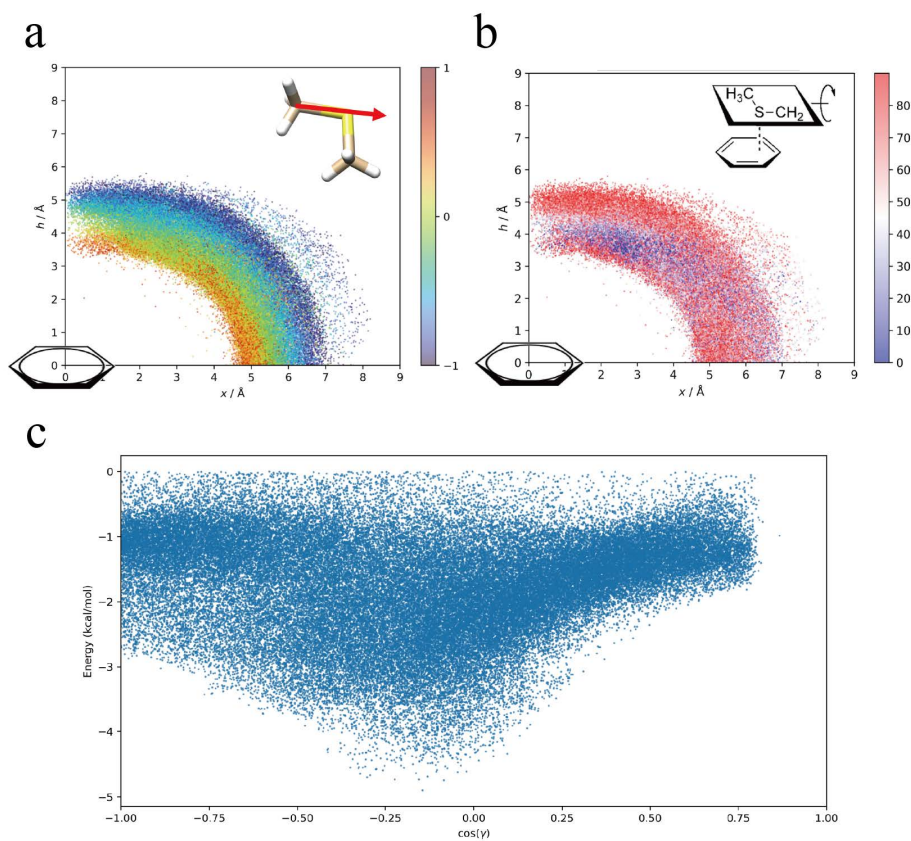

**Figure S7.** Radial distribution of MSM around aromatic rings, with *x* as the centroid shift (Å) and *h* as the height above the plane (Å), with point colors indicating the trend of the geometric parameter of interest. (a) Orientation distribution of the C→S vector. The color bar denotes orientation (cos$\gamma$): values > 0 (red) correspond to the C→S vector pointing toward the ring center, whereas values < 0 (blue) indicate the opposite direction. The inset (upper right) is the schematic of the C–S bond definition. (b) Interplanar angle distribution between the plane of the MSM fragment and the aromatic ring plane. The inset (upper right) illustrates the definition of the MSM fragment plane used in the analysis. (c) Predicted interaction energies (kcal/mol) for MBZ-MSM as a function of the C–S bond orientation (cos$\gamma$, where 1 = toward the aromatic ring, –1 = away, 0 = parallel).

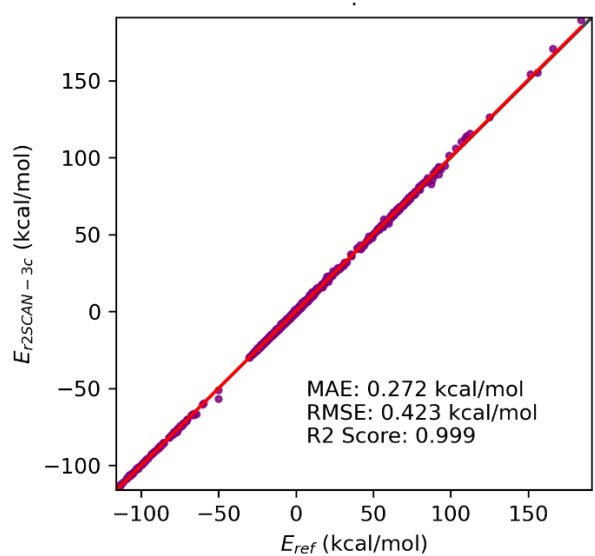

**Figure S8.** Comparison of interaction energies calculated at the ωB97X-D3BJ/def2-TZVPP level and r²SCAN-3c for randomly selected 30 dimers from each class.

# 7 NCI Visualization

The Reduced Density Gradient (RDG) method[4] is used to visualize weak interactions. In this study, weak interaction analysis[5] was performed using the Multiwfn program[6], and the corresponding visualizations were generated with VMD software[7]. In the weak interaction analysis, the gradient isosurfaces of the RDG function are typically colored on a scale ranging from blue-green to red, corresponding to interaction strength from strong attraction to strong repulsion. Blue regions indicate strong attractive interactions, green regions correspond to weak van der Waals interactions, and red regions denote strong steric repulsion.

# 8 Supporting Information Tables

**Table S1.** Data volume of each fragment

| Fragment type | Total number* | Training number**(%)*** |
|---|---|---|
| ACEM | 4,830,133 | 533,043 (11.03) |
| ACET | 5,246,390 | 1,170,051 (22.30) |
| ETAM | 1,480,001 | 347,779 (23.50) |
| ETOH | 3,299,189 | 374,098 (11.34) |
| ETSH | 350,411 | 79,419 (22.66) |
| MBZ | 3,380,598 | 252,517 (7.47) |
| MGDM | 2,520,733 | 474,908 (18.84) |
| Imidazole | 1,377,566 | 284,607 (20.66) |
| MIND | 1,325,598 | 194,941 (14.71) |
| PMPO | 1,697,412 | 258,379 (15.22) |
| MSM | 949,756 | 141,498 (14.90) |
| N1PA | 1,103,276 | 179,516 (16.27) |
| NMA | 22,026,093 | 876,434 (3.98) |
| PRPA | 9,306,132 | 527,630 (5.67) |
| HOH | 3,561,146 | 327,658 (9.20) |

*The total number of occurrences in the original dataset.

**The number of occurrences in the training set.

***Percentage of total data used for training.

**Table S2.** Calculation time of each fragment

| Fragment Type | Time* (in hours) | |
|---|---|---|
| | ωB97X-D3BJ/def2-TZVPP | PANIP |
| ACEM | 36 days, 20:03:14.2 | 0:39:18.0 |
| ACET | 20 days, 9:32:51.3 | 0:37:32.9 |
| ETAM | 45 days, 11:24:32.5 | 0:40:42.3 |
| ETOH | 34 days, 18:38:02.8 | 0:39:19.3 |
| ETSH | 29 days, 0:16:31.1 | 0:39:03.9 |
| MBZ | 53 days, 17:01:56.3 | 0:44:10.8 |
| MGDM | 66 days, 5:37:29.9 | 0:41:39.7 |
| MIMD | 56 days, 10:43:06.7 | 0:40:31.7 |
| MIME | 54 days, 12:43:22.4 | 0:41:42.3 |
| MIMM | 58 days, 13:51:29.8 | 0:40:60.0 |
| MIND | 108 days, 13:58:41.7 | 0:46:57.3 |
| PMPO | 80 days, 17:58:59.7 | 0:44:01.9 |
| MSM | 35 days, 0:20:00.7 | 0:39:58.2 |
| N1PA | 94 days, 2:16:00.9 | 0:46:23.8 |
| NMA | 47 days, 5:44:10.5 | 0:42:11.5 |
| PRPA | 36 days, 13:47:55.2 | 0:41:15.4 |
| HOH | 12 days, 21:28:45.8 | 0:35:22.9 |
| Total | 463 days, 11:19:06.1 | 6:17:55.6 |

*All benchmark calculations were performed on the computer cluster with an Intel Xeon Platinum 8368 CPU @ 2.60 GHz with 504 GB RAM.